\documentclass[prb,twocolumn,showpacs,floatfix]{revtex4}

\usepackage{amsmath,amssymb,graphicx}

\DeclareMathOperator{\re}{Re}
\DeclareMathOperator{\erfc}{erfc}

\begin{document}

\title{Edge states in bilayer graphene in a magnetic field}

\author{P. K. Pyatkovskiy}
\affiliation{Department of Applied Mathematics, Western University, London, Ontario N6A 5B7, Canada}

\begin{abstract}
Edge states in biased bilayer graphene in a magnetic field are studied within the four-band continuum model. The analysis is done for the semi-infinite graphene plane and for the graphene ribbon of a finite width, in the cases of zigzag and armchair edges. Exact dispersion equations for the edge states and analytic expressions for their wave functions are written in terms of the parabolic cylinder functions. The spectrum of edge states for each type of the boundary conditions is found by numerically solving the corresponding dispersion equations. The low-energy modes localized at zigzag edges are explored in detail.
\end{abstract}

\pacs{73.22.Pr, 73.43.--f, 71.70.Di}

\maketitle

\section{Introduction}

Transport properties of a two-dimensional system can be significantly affected by the presence of edge states. These quasi-one-dimensional states, localized at the boundary of the sample, may provide the current-carrying channels even when the bulk excitations are gapped. In a magnetic field, edge states at the Fermi level give an important contribution to the Hall conductance of a two-dimensional electron gas.\cite{Halperin1982PRB}

The edge state spectrum in graphene, an atomically thin layer of carbon atoms arranged in a honeycomb crystalline lattice,~\cite{Novoselov2004S} depends on the type of the edge termination. There are two typical shapes of a graphene edge, \emph{zigzag} and \emph{armchair}. In addition to the quantum Hall edge states, zigzag edges of monolayer graphene support the dispersionless zero-energy edge states,\cite{Peres2006PRB,Brey2006PRB,Abanin2006PRL} present even without magnetic field.\cite{Fujita1996JPSJ,Nakada1996PRB} The spin splitting of the lowest Landau level results in the counterpropagating quantum Hall edge states with opposite spin polarization~\cite{Abanin2006PRL,Fertig2006PRL,Abanin2007SSC} at zero chemical potential ($\nu=0$ state). In a more general case of the quantum Hall ferromagnetic order parameters\cite{Nomura2006PRL,Goerbig2006,Alicea2006PRB} and the magnetic catalysis parameters (Dirac masses),\cite{Gusynin2006catalysis,Herbut2006,Ezawa2006} the existence of the gapless edge states depends both on the ratio of the different order parameters and the edge type.\cite{Gusynin2008PRB,Gusynin2009PRB}

Bilayer graphene consists of two $AB$ (Bernal) stacked graphene monolayers. The spectrum gap in this system can be tuned by applying a gate voltage (bias) which creates the charge imbalance between the two layers.\cite{McCann2006PRL,Ohta2006S} In the case of zigzag edges, in addition to the dispersionless modes similar to the ones that exist in monolayer graphene, there are also dispersive subgap edge excitations that carry counterpropagating currents in two valleys at a given edge.\cite{Castro2008PRL,Li2010PRB,Li2011NP} In the presence of a perpendicular magnetic field, the charge imbalance between the two layers leads to the valley splitting of the zero-energy Landau levels, which manifests itself in experiments as an extra $\nu=0$ quantum Hall plateau.\cite{Castro2007PRL} The edge state structure in this regime in the case of zigzag edges has been studied both by the tight-binding method\cite{Castro2007PRL,Mazo2011PRB,Wu2012PRB,Zhang2012PRB} and by the perturbation and variational methods within the continuum (Dirac) model.\cite{Mazo2011PRB} The calculations of the edge state spectrum in bilayer graphene with armchair edges in a magnetic field have so far been limited to narrow samples where the Landau level formation occurs only at unrealistic field magnitudes.\cite{Xu2009PRB} The aim of the present paper is to study the edge state spectrum of a bilayer graphene ribbon or a semi-infinite plane with zigzag or armchair edges in a magnetic field using the exact solutions to the differential equations of the continuum model, by combining analytic and numerical methods.

The low-energy edge state spectrum is found to be qualitatively different for the two edge types. In the case of zigzag edges, two zero-energy states per edge and spin are present at all accessible magnetic field values, which is consistent with previous findings.\cite{Castro2007PRL,Mazo2011PRB,Wu2012PRB,Zhang2012PRB} Furthermore, one of these states is shown to be almost independent of a magnetic field strength, whereas the other one exhibits the partial hybridization with the bulk state $n=1$. In contrast, in the case of armchair edges the spectrum is gapped and zero-energy states are absent.

The paper is organized as follows. In Sec.~\ref{secII} we introduce the four-band continuum model for bilayer graphene in a magnetic field and present the general solution for the wave functions in the translationally invariant along the $x$ axis case. The dispersion equations for edge states are derived and solved in Sec.~\ref{secIII} in the cases of zigzag edges and in Sec.~\ref{secIV} in the case of armchair edges. In Sec.~\ref{secV} we give a brief summary of our results. Detailed derivations of the general solution and its different asymptotes are given in three Appendices.

\section{General solution in the continuum model}
\label{secII}

\subsection{Model}

\begin{figure}
\includegraphics[width=0.65\columnwidth]{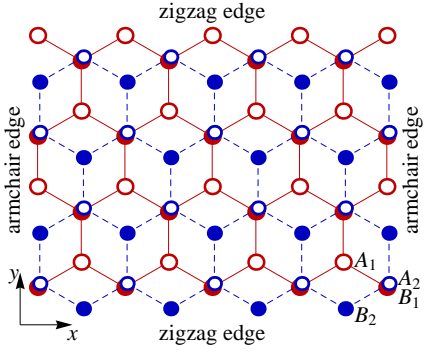}
\caption{The lattice structure of bilayer graphene with zigzag and armchair edges.}
\label{lattice}
\end{figure}

We consider  bilayer graphene with Bernal stacking (Fig.~\ref{lattice}), taking into account only the nearest-neighbor in-plane hopping $t\simeq3$\,eV and the interlayer $A_2B_1$ hopping $\gamma_1\simeq0.4$\,eV. We limit ourselves to the case of perfect zigzag or armchair edges, neglecting the effects of disorder and electron-electron interactions. The effective four-band Hamiltonian for noninteracting electrons in each valley $K_{\xi=\pm}$ is\cite{McCann2006PRL}
\begin{equation}
H_\xi=
\xi
\begin{pmatrix}
\Delta & v_F\hat\pi^\dag & 0 & 0 \\
v_F\hat\pi & \Delta & \xi\gamma_1 & 0 \\
0 & \xi\gamma_1 & -\Delta & v_F\hat\pi^\dag \\
0 & 0 & v_F\hat\pi & -\Delta
\end{pmatrix}
,
\label{blg_ham}
\end{equation}
where $\hat\pi=\hat p_x+i\hat p_y$, $v_F=\sqrt3ta/(2\hbar)\simeq10^6$\,m/s is the Fermi velocity, and $a\simeq0.246$\,nm is the lattice constant of graphene. The magnitude of the layer asymmetry gap parameter $\Delta=U/2$, where $U$ is the interlayer potential induced by the applied perpendicular electric field, is bound by the relation $2|\Delta|<\gamma_1$.~\cite{Zhang2010PRB} The external magnetic field $\mathbf B=[\nabla\times\mathbf A]=(0,0,B)$ is perpendicular to the graphene plane (we assume $B>0$), and the momentum operator is $\hat{\mathbf p}=-i\hbar\nabla+(e/c)\mathbf A$ with the electron charge  $-e<0$.

The Hamiltonian~(\ref{blg_ham}) acts on wave functions, the components of which correspond to different layers and sublattices. They are written in valleys $K_+$ and $K_-$ as
\begin{equation}
\Psi_{+}^s=
\begin{pmatrix}
\Psi_{+A_1}^s \\
\Psi_{+B_1}^s \\
\Psi_{+A_2}^s \\
\Psi_{+B_2}^s
\end{pmatrix}
,
\qquad
\Psi_{-}^s=
\begin{pmatrix}
\Psi_{-B_2}^s \\
\Psi_{-A_2}^s \\
\Psi_{-B_1}^s \\
\Psi_{-A_1}^s
\end{pmatrix}
,
\label{components}
\end{equation}
where $s=\pm$ is the additional spin index.

\subsection{General solution with \textit{x} translational invariance}

For edges along the $x$ axis (this orientation corresponds to the zigzag edge type; see Fig.~\ref{lattice}), it is suitable to use the Landau gauge $\mathbf A=(-By,0)$. The wave functions are plane waves in the $x$ direction,
\begin{equation}
\Psi^s_\xi(x,y)=e^{ikx}\Psi^s_\xi(y,k),
\end{equation}
where the envelope functions $\Psi^s_\xi(y,k)\equiv\Psi^s_\xi(\eta)$ depend only on a single combination of the variables, $\eta=y/l-kl$, with $l=\sqrt{\hbar c/(eB)}$ being the magnetic length. They satisfy the equation
\begin{equation}
\xi
\begin{pmatrix}
\Delta & -\epsilon_0\hat a & 0 & 0 \\
-\epsilon_0\hat a^\dag & \Delta & \xi\gamma_1 & 0 \\
0 & \xi\gamma_1 & -\Delta & -\epsilon_0\hat a \\
0 & 0 & -\epsilon_0\hat a^\dag & -\Delta
\end{pmatrix}
\Psi^s_\xi(\eta)=E\Psi^s_\xi(\eta),
\label{E_eig_eq}
\end{equation}
where $\hat a=2^{-1/2}(\eta+\partial_\eta)$ and $\hat a^\dag=2^{-1/2}(\eta-\partial_\eta)$ are the annihilation and creation operators and $\epsilon_0=\sqrt2\hbar v_F/l\simeq36\sqrt{B[\mathrm T]}$\,meV is the cyclotron energy in monolayer graphene.

The general solution of the system of differential equations~(\ref{E_eig_eq}) is the linear combination
\begin{equation}
\Psi^s_\xi(\eta)=\sum_{i=1}^4C_i^\xi\Phi_\xi^{(i)}(\eta)
\end{equation}
of four independent functions (see Appendix~\ref{Appendix:solution} for details),
\begin{align}
\Phi_\xi^{(1)}(\eta) &= f^\xi_{\lambda_1^\xi}(\eta), \nonumber \\
\Phi_\xi^{(2)}(\eta) &=
\bigl[f^\xi_{\lambda_1^\xi}(\eta)-f^\xi_{\lambda_2^\xi}(\eta)\bigr]/(\lambda_1^\xi-\lambda_2^\xi), \nonumber \\
\Phi_\xi^{(3)}(\eta) &= h^\xi_{\lambda_1^\xi}(\eta), \nonumber \\
\Phi_\xi^{(4)}(\eta) &=
\bigl[h^\xi_{\lambda_1^\xi}(\eta)-h^\xi_{\lambda_2^\xi}(\eta)\bigr]/(\lambda_1^\xi-\lambda_2^\xi),
\label{gen_sol}
\end{align}
where
\begin{align}
\lambda_{1,2}^\xi &= \frac12+\frac{E^2+\Delta^2}{\epsilon_0^2} \nonumber \\
&\quad\pm\frac{\sqrt{(\epsilon_0^2-4\xi\Delta E)^2+4\gamma_1^2(E^2-\Delta^2)}}{2\epsilon_0^2},
\label{lambdas}
\end{align}
are two energy-dependent dimensionless parameters (in general, complex) and the individual solutions $f^\xi_{\lambda_i^\xi}(\eta)$ and $h^\xi_{\lambda_i^\xi}(\eta)$ are written in terms of the parabolic cylinder functions $U(a,z)$ and $V(a,z)$,~\cite{Abramowitz}
\begin{align}
f^\pm_\lambda(\eta) &=
\begin{pmatrix}
\pm\nu_\pm(\lambda)U\bigl(\frac32-\lambda,\sqrt2\eta\bigr) \\
\frac{(E\pm\Delta)^2-\epsilon_0^2\lambda}{\epsilon_0\gamma_1}U\bigl(\frac12-\lambda,\sqrt2\eta\bigr) \\
\frac{E\pm\Delta}{\epsilon_0}U\bigl(\frac12-\lambda,\sqrt2\eta\bigr) \\
\mp U\bigl(-\frac12-\lambda,\sqrt2\eta\bigr)
\end{pmatrix}
,\label{f}
\\
h^\pm_\lambda(\eta) &=
\begin{pmatrix}
\mp V\bigl(\frac32-\lambda,\sqrt2\eta\bigr) \\
\frac{E\mp\Delta}{\epsilon_0}V\bigl(\frac12-\lambda,\sqrt2\eta\bigr) \\
\frac{(E\mp\Delta)^2-\epsilon_0^2(\lambda-1)}{\epsilon_0\gamma_1}V\bigl(\frac12-\lambda,\sqrt2\eta\bigr) \\
\pm\nu_\mp(\lambda-1)V\bigl(-\frac12-\lambda,\sqrt2\eta\bigr)
\end{pmatrix}
,
\label{h}
\end{align}
with
\begin{equation}
\nu_\pm(\lambda)=\frac{(E\pm\Delta)(\gamma_1^2+\Delta^2-E^2)+(E\mp\Delta)\epsilon_0^2\lambda}{\epsilon_0^2\gamma_1}.
\end{equation}

\subsection{Bulk solutions}

On an infinite plane, the normalizable wave functions contain only the parabolic cylinder functions $U(a,z)$ which are bounded at $z\to\pm\infty$ provided that $a=-n-1/2$, where $n$ is a nonnegative integer. In this case, the following relation is valid:
\begin{equation}
U(-n-1/2,\sqrt2\eta)=2^{-\frac n2}e^{-\frac{\eta^2}2}H_n(\eta),\quad n=0,1,2,\dots,
\end{equation}
where $H_n(z)$ are the Hermite polynomials. Therefore, there is a nontrivial bounded solution proportional to $f^\xi_n(\eta)$ on an infinite plane when $\lambda_1^\xi=n$ or $\lambda_2^\xi=n$ with $n=2,3,4,\dots$. This condition is equivalent to the quartic equation for the energy of bulk Landau levels~\cite{Pereira2007PRB}
\begin{align}
\bigl[(E+\xi\Delta)^2-n\epsilon_0^2\bigr]\bigl[(E-\xi\Delta&)^2-(n-1)\epsilon_0^2\bigr] \nonumber \\
&-\gamma_1^2(E^2-\Delta^2)=0.
\label{bulk_LL_eq}
\end{align}
For each $\xi=\pm$ and $n=2,3,4,\dots$, it has four solutions $E_{\pm n}^{\kappa,\xi}$, where $\kappa=+(-)$ corresponds to the high (low) energy band. All high-energy band Landau levels have energies satisfying $(E_{\pm n}^{+,\xi})^2>\gamma_1^2+\Delta^2$.

As seen from Eq.~(\ref{f}), at $\lambda_i^\xi=1$ the solution proportional to $f^\xi_1(\eta)$  is normalizable on an infinite plane provided that $\nu_\xi(1)=0$. Therefore, in addition to the solutions of Eq.~(\ref{bulk_LL_eq}), there are two more high-energy levels $E_{\pm1}^{+,\xi}$ and one low-energy level $E_1^{-,\xi}$, given by the roots of the cubic equation
\begin{equation}
\label{bulk_LL_eq_n1}
(E+\xi\Delta)(\gamma_1^2+\Delta^2-E^2)+(E-\xi\Delta)\epsilon_0^2=0.
\end{equation}
Finally, in the case $\lambda^\xi_i=0$ there is a valid solution proportional to $f^\xi_0(\eta)$ on an infinite plane provided that $\nu_\xi(0)=0$ and $E+\xi\Delta=0$, which yields the remaining low-energy level
\begin{equation}
E_0^{-,\xi}=-\xi\Delta.
\label{bulk_LL_n0}
\end{equation}

The corresponding bulk wave functions are~\cite{Pereira2007PRB,Nakamura2009PRL}
\begin{equation}
\begin{split}
&\Psi^s_\xi(\eta) =
C^\xi_1f_n^\xi(\eta) \\
&=C^\xi_12^{\frac{2-n}2}e^{-\frac{\eta^2}2}
\begin{pmatrix}
\frac{(n-1)[n\epsilon_0^2-(E_{\pm n}^{\kappa,\xi}+\xi\Delta)^2]}
{\gamma_1(\xi E_{\pm n}^{\kappa,\xi}-\Delta)}
H_{n-2}(\eta)
\\
\frac{(E_{\pm n}^{\kappa,\xi}+\xi\Delta)^2-n\epsilon_0^2}{\sqrt2\epsilon_0\gamma_1}
H_{n-1}(\eta)
\\
\frac{E_{\pm n}^{\kappa,\xi}+\xi\Delta}{\sqrt2\epsilon_0}H_{n-1}(\eta)
\\
-\frac\xi2H_n(\eta)
\end{pmatrix}
.
\label{sol_inf_plane}
\end{split}
\end{equation}

In the case of unbiased bilayer graphene ($\Delta=0$), the Landau level energies are equal in both valleys $K_\pm$ and are given by expression~\cite{Pereira2007PRB}
\begin{align}
E^{\kappa,\xi}_{\pm n} &=
\pm\frac1{\sqrt2}\Bigl(\bigl|\gamma_1^2+(2n-1)\epsilon_0^2\bigr| \nonumber \\
&\qquad+\kappa\sqrt{(\gamma_1^2-\epsilon_0^2)^2+4n\gamma_1^2\epsilon_0^2}\Bigr)^{1/2}.
\end{align}
Finite $\Delta$ causes the valley splitting of Landau levels as well as the splitting between levels $n=0$ and $n=1$.~\cite{Nakamura2009PRL}

In the case $\gamma_1\gg\epsilon_0\gg|\Delta|$, the low-energy Landau levels are approximately given by the two-band effective model~\cite{McCann2006PRL}
\begin{align}
E^{-,\xi}_{\pm n}&\simeq\pm\sqrt{\hbar^2\omega_c^2n(n-1)+\Delta^2}, \quad n=2,3,4,\dots , \\
E^{-,\xi}_1&\simeq E^{-,\xi}_0=-\xi\Delta,
\end{align}
where $\omega_c$ is the cyclotron frequency in bilayer graphene, $\hbar\omega_c=\epsilon_0^2/\gamma_1\simeq3.2B[\mathrm T]$\,meV.

\section{Zigzag edges}
\label{secIII}

\subsection{Dispersion equations for half plane}

On a semi-infinite plane $y>0$, the normalizable wave functions are given in terms of only $U(a,z)$ function, which decreases exponentially as $z\to\infty$, while the function $V(a,z)$ grows exponentially in both directions $z\to\pm\infty$. Therefore, $C^\xi_3=C^\xi_4=0$ and the solution is
\begin{equation}
\Psi^s_\xi(\eta)=C_1^\xi\Phi_\xi^{(1)}(\eta)+C_2^\xi\Phi_\xi^{(2)}(\eta).
\label{gen_sol_hp}
\end{equation}
In the limit $\Delta\to0$, this solution reduces to the one used in Ref.~\onlinecite{Koshino2010PRB} in the description of interface states on the monolayer-bilayer graphene junction.

The boundary conditions at the zigzag edge $y=0$ of the half plane (which corresponds to $\eta=-kl$) are~\cite{Mazo2011PRB}
\begin{equation}
\Psi^s_{\xi A_i}(-kl)=0,\qquad i=1,2;
\label{zigzag_half_plane_BC}
\end{equation}
i.e., the wave functions on $A$ atoms should vanish at the edge on both layers. These boundary  conditions do not mix the components of the wave functions from different valleys ($\xi=\pm$) and lead to the following system of two equations for each valley $K_\xi$:
\begin{equation}
C_1^\xi\Phi_{\xi A_i}^{(1)}(-kl)+C_2^\xi\Phi_{\xi A_i}^{(2)}(-kl)=0, \qquad i=1,2,
\end{equation}
where the layer and sublattice components of $\Phi^{(j)}_\xi$, $f^\xi_\lambda$, and $h^\xi_\lambda$ are chosen in the same way as the components of $\Psi^s_\xi$ in Eq.~(\ref{components}).
These systems have nontrivial solutions when the corresponding determinants of the coefficient functions are zero; i.e.,
\begin{equation}
\begin{vmatrix}
\Phi_{\xi A_1}^{(1)}(-kl) & \Phi_{\xi A_1}^{(2)}(-kl) \\
\Phi_{\xi A_2}^{(1)}(-kl) & \Phi_{\xi A_2}^{(2)}(-kl)
\end{vmatrix}
=0,
\end{equation}
which is equivalent to
\begin{equation}
\frac1{\lambda^\xi_2-\lambda^\xi_1}
\begin{vmatrix}
f^{\xi A_1}_{\lambda_1^\xi}(-kl) & f^{\xi A_1}_{\lambda_2^\xi}(-kl) \\
f^{\xi A_2}_{\lambda_1^\xi}(-kl) & f^{\xi A_2}_{\lambda_2^\xi}(-kl)
\end{vmatrix}
=0.
\end{equation}
Writing the components of $f^\xi_{\lambda^\xi_i}(-kl)$ explicitly, one arrives at the dispersion equation for the $K_+$ valley,
\begin{equation}
\begin{split}
&\frac{E+\Delta}{\lambda^+_2-\lambda^+_1}
\biggl\{\nu_+(\lambda^+_1)U\biggl(\frac32-\lambda^+_1,-\sqrt2kl\biggr) \\
&\quad\times U\biggl(\frac12-\lambda^+_2,-\sqrt2kl\biggr)
-\bigl(\lambda^+_1\leftrightarrow\lambda^+_2\bigr)\biggr\}=0,
\label{hp_zigzag_disp_eq_plus}
\end{split}
\end{equation}
and the $K_-$ valley,
\begin{equation}
\begin{split}
&\frac1{\lambda^-_2-\lambda^-_1}
\biggl\{\bigl[(E-\Delta)^2-\epsilon_0^2\lambda_1^-\bigr]U\biggl(\frac12-\lambda_1^-,-\sqrt2kl\biggr) \\
&\quad\times U\biggl(-\frac12-\lambda_2^-,-\sqrt2kl\biggr)
-\bigl(\lambda^-_1\leftrightarrow\lambda^-_2\bigr)\biggr\}=0.
\label{hp_zigzag_disp_eq_minus}
\end{split}
\end{equation}

\subsection{Dispersion equations for ribbon}

The boundary conditions at two ribbon edges $y=0$ and $y=W$ (corresponding to $\eta=-kl$ and $\eta=W/l-kl$, respectively) are~\cite{Mazo2011PRB}
\begin{equation}
\psi^s_{\xi A_i}(-kl)=\Psi^s_{\xi B_i}(W/l-kl)=0,\qquad i=1,2;
\end{equation}
i.e., in addition to the condition~(\ref{zigzag_half_plane_BC}) at the edge $y=0$, the components $B_{1,2}$ must vanish at the opposite edge. These boundary  conditions also do not mix valleys and imply the following independent system of four equations for each valley:
\begin{equation}
\begin{array}{l}
\sum_{j=1}^4C_j^\xi\Phi_{\xi A_i}^{(j)}(-kl)=0, \smallskip
\\
\sum_{j=1}^4C_j^\xi\Phi_{\xi B_i}^{(j)}(W/l-kl)=0, \qquad i=1,2.
\end{array}
\end{equation}
These systems have nontrivial solutions when the corresponding determinants of coefficient functions are zero. After some straightforward algebra, we obtain the dispersion equation for the $K_+$ valley,
\begin{equation}
\frac1{(\lambda_2^+-\lambda_1^+)^2}
\begin{vmatrix}
X_1^+(-kl) & Y_1^+(W/l-kl) \\
X_2^+(-kl) & Y_2^+(W/l-kl)
\end{vmatrix}
=0,
\label{r_zigzag_disp_eq_plus}
\end{equation}
and the $K_-$ valley,
\begin{equation}
\frac1{(\lambda_2^--\lambda_1^-)^2}
\begin{vmatrix}
X_1^-(W/l-kl) & Y_1^-(-kl) \\
X_2^-(W/l-kl) & Y_2^-(-kl)
\end{vmatrix}
=0,
\label{r_zigzag_disp_eq_minus}
\end{equation}
where the $2\times2$ blocks $X_i^\pm(\eta)$ and $Y_i^\pm(\eta)$ are defined as
\begin{widetext}
\begin{align}
X_i^\pm(\eta) &=
\begin{pmatrix}
\nu_\pm(\lambda^\pm_i)
U\bigl(\frac32-\lambda^\pm_i,\sqrt2\eta\bigr) &
\frac{E\pm\Delta}{\epsilon_0}U\bigl(\frac12-\lambda^\pm_i,\sqrt2\eta\bigr) \\
-V\bigl(\frac32-\lambda^\pm_i,\sqrt2\eta\bigr) &
\frac{(E\mp\Delta)^2-\epsilon_0^2(\lambda^\pm_i-1)}{\epsilon_0\gamma_1}
V\bigl(\frac12-\lambda^\pm_i,\sqrt2\eta\bigr)
\end{pmatrix},
\label{X}
\\
Y_i^\pm(\eta) &=
\begin{pmatrix}
-U\bigl(-\frac12-\lambda^\pm_i,\sqrt2\eta\bigr) &
\frac{(E\pm\Delta)^2-\epsilon_0^2\lambda^\pm_i}{\epsilon_0\gamma_1}
U\bigl(\frac12-\lambda^\pm_i,\sqrt2\eta\bigr) \\
\nu_\mp(\lambda^\pm_i-1)
V\bigl(-\frac12-\lambda^\pm_i,\sqrt2\eta\bigr) &
\frac{E\mp\Delta}{\epsilon_0}V\bigl(\frac12-\lambda^\pm_i,\sqrt2\eta\bigr)
\end{pmatrix}.
\label{Y}
\end{align}
\end{widetext}
Using the expressions for $U(a,z)$ and $V(a,z)$ in terms of the parabolic cylinder function $D_\lambda(z)$,
\begin{equation}
U\biggl(-\frac12-\lambda,z\biggr)=D_\lambda(z),
\label{UDrelation}
\end{equation}
\begin{equation}
V\biggl(-\frac12-\lambda,z\biggr)=\frac{\Gamma(-\lambda)}\pi
\bigl[D_\lambda(-z)-\cos(\pi\lambda)D_\lambda(z)\bigr],
\label{VDrelation}
\end{equation}
one can show the following symmetry property of the determinants in Eqs.~(\ref{r_zigzag_disp_eq_plus}) and~(\ref{r_zigzag_disp_eq_minus}):
\begin{equation}
\begin{vmatrix}
X_1^\xi(\eta_1) & Y_1^\xi(\eta_2) \\
X_2^\xi(\eta_1) & Y_2^\xi(\eta_2)
\end{vmatrix}
=
\begin{vmatrix}
X_1^\xi(-\eta_1) & Y_1^\xi(-\eta_2) \\
X_2^\xi(-\eta_1) & Y_2^\xi(-\eta_2)
\end{vmatrix}
.
\end{equation}
It implies that the energy spectra in two valleys are related by
\begin{equation}
E^\xi(k)=-E^{-\xi}(W/l^2-k)
\label{zigzag_E_symmetry}
\end{equation}
(note that the momenta in each valley are measured from the corresponding $K$ points).
The corresponding wave functions are related by
\begin{equation}
\Psi^s_\xi(y,k)=C
\begin{pmatrix}
\sigma_3 & 0  \\
0 & \sigma_3
\end{pmatrix}
\Psi^s_{-\xi}(W-y,W/l^2-k),
\label{zigzag_psi_symmetry}
\end{equation}
where $\sigma_3$ is the Pauli matrix.

\subsection{Results for the spectra}

We numerically solve dispersion equations~(\ref{hp_zigzag_disp_eq_plus}) and~(\ref{hp_zigzag_disp_eq_minus}) in the case of the semi-infinite plane and Eqs.~(\ref{r_zigzag_disp_eq_plus}) and~(\ref{r_zigzag_disp_eq_minus}) in the case of the finite-width ribbon. The solutions include both low-energy band ($\kappa=-1$) and high-energy band ($\kappa=+1$) spectrum branches. In what follows, we limit our consideration to energies lower than $\gamma_1$ and focus only on the low-energy branches. It is also assumed that $\Delta\geqslant0$, taking into account that in the zigzag edge case the change of the sign of $\Delta$ results in merely the inversion $E\to-E$ of the spectrum.

\begin{figure*}
\includegraphics[width=\textwidth]{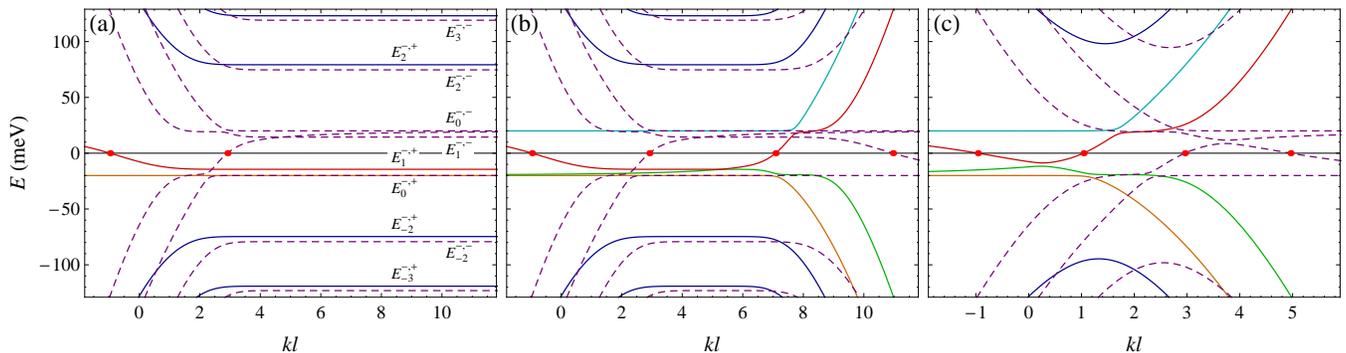}
\caption{Numerical results for the low-energy spectrum in bilayer graphene with zigzag edge(s) at $B=20\,\mathrm T$ and $\Delta=20\,\mathrm{meV}$: (a)~half plane, (b)~ribbon of the width $W=10l$, (c)~ribbon of the width $W=4l$. Solid (dashed) lines represent the spectrum in the $K_+$ ($K_-$) valley. In panel (a), the bulk Landau level energies $E_n^{\kappa,\xi}$ are indicated. Gapless edge states at $\mu_s=0$ are marked by dots.}
\label{zigzag_fig}
\end{figure*}

First, we consider the case $2\Delta<\epsilon_0$ when $n=1$ and $n=2$ are the two Landau levels with the lowest energies. Our results are consistent with those obtained previously in the tight-binding studies.\cite{Castro2007PRL,Mazo2011PRB,Wu2012PRB,Zhang2012PRB} The examples of the spectra showing a few lower Landau levels in the case of a half plane and two different widths of the ribbon are shown in Fig.~\ref{zigzag_fig}. On a half plane, the spectrum branches at $kl\gg1$ asymptotically approach the bulk Landau levels given by Eqs.~(\ref{bulk_LL_eq})--(\ref{bulk_LL_n0}). The states corresponding to these asymptotes (plateaus) are approximately described by infinite-plane solutions~(\ref{sol_inf_plane}) with $\eta=y/l-kl$. They are localized in the bulk and centered along the $y$ direction at $y_k=kl^2$ (the position wave-vector duality in Landau gauge). The same is true for wide ribbons $W\gg l$ [see Fig.~\ref{zigzag_fig}(b)], where the plateaus closely approaching the bulk Landau levels are formed. For a given branch, all states to the left (right) of the bulk plateau are localized in the vicinity of the edge $y=0$ ($y=W$). There are also two purely edge state branches in each valley, which do not correspond to any of the bulk Landau levels. The states on these branches, as well as the edge states corresponding to the levels $n=0,1$, remain present even at $B=0$. As we will see below, the main effect of magnetic field on these edge state modes is the relative horizontal shift $\delta k=W/l^2$ between the states on the opposite edges.

The width of a given bulk plateau is determined by the range of $y_k$ for which the corresponding bulk wave function~(\ref{sol_inf_plane}) remains almost unperturbed by the edges. Due to the increase of the localization length of the bulk state with increasing $|n|$, the widths of the higher bulk Landau level plateaus are smaller. In the case of a narrow ribbon, shown in Fig.~\ref{zigzag_fig}(c), the bulk Landau level plateaus are not formed.

\begin{figure}
\includegraphics[width=0.95\columnwidth]{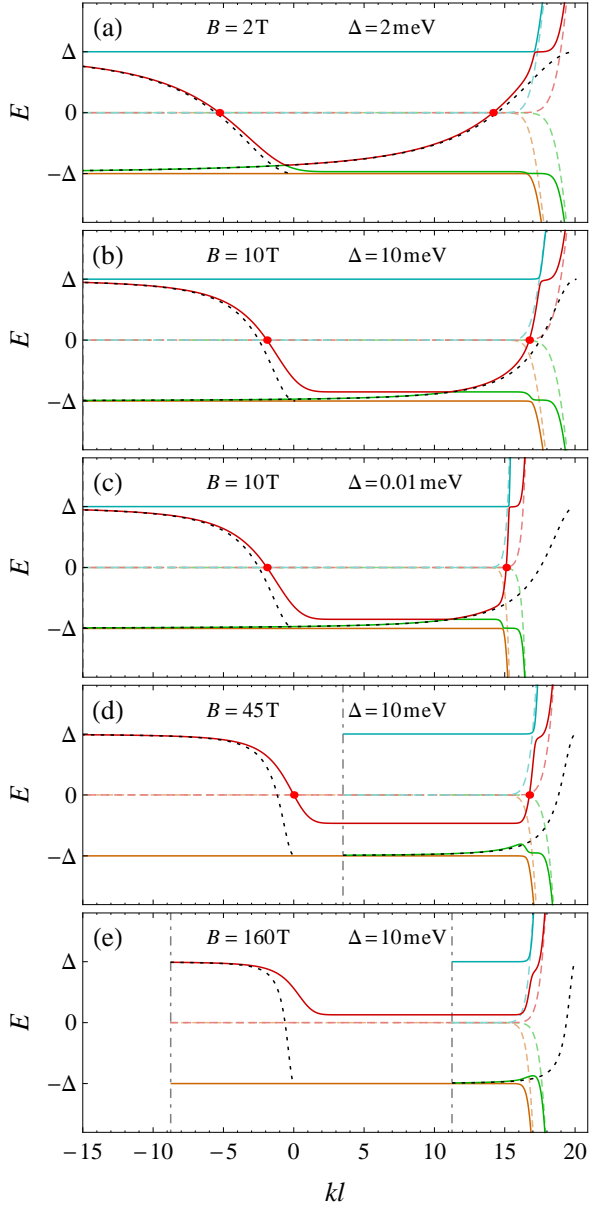}
\caption{Numerical results for the lowest spectrum branches in the $K_+$ valley for zigzag ribbon of a width $W=20l$ at different values of magnetic field and gap parameter $\Delta\ll\epsilon_0$. Dashed lines correspond to the unbiased case ($\Delta=0$), and dotted lines show the subgap edge modes at $B=0$ (right edge mode is shifted horizontally with $k\to k+W/l^2$ in order to illustrate the effect of a magnetic field). Dot-dashed vertical lines display the cutoffs for the edge modes. Gapless edge states at $\mu_s=0$ are marked by dots.}
\label{LLLs}
\end{figure}

\begin{figure}
\includegraphics[width=0.95\columnwidth]{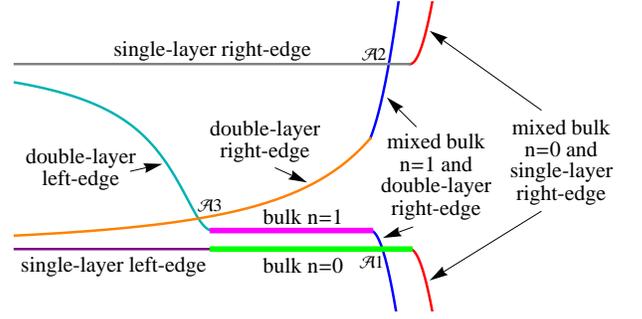}
\caption{Schematic illustration of the spectrum branches with the lowest energies in the $K_+$ valley for wide zigzag ribbon ($W\gg l$) at small gap parameter and moderate magnetic field ($\Delta\ll\epsilon_0\ll\gamma_1$), indicating the properties of the corresponding states. Level splittings at avoided crossings $\mathcal A1$, $\mathcal A2$, $\mathcal A3$ of the branches are removed for clarity. Bulk Landau levels are shown by thick lines.}
\label{schematic}
\end{figure}

In the following, we consider only the spectrum in the $K_+$ valley, taking into account that the energies and the wave functions in the two valleys are related by Eqs.~(\ref{zigzag_E_symmetry}) and~(\ref{zigzag_psi_symmetry}). The structure of the spectrum at the energy scale $|E|\lesssim\Delta$ is shown in Fig.~\ref{LLLs} for different values of a magnetic field and $\Delta$, and the properties of the corresponding states are given in Fig.~\ref{schematic}. This low-energy spectrum consists of the four branches and is complicated by their avoided crossings. For the moment, we ignore the level splittings at these anticrossings and briefly describe the eigenstates corresponding to each branch. We assume that the ribbon is wide enough ($W/l\gg1$) so that the plateaus corresponding to bulk states $n=0$ and $n=1$ are formed, and in the middle of these plateaus the effects of the edges on the bulk wave functions can be neglected. 

The wave function~(\ref{sol_inf_plane}) of the $n=0$ bulk Landau level resides solely on a single layer and sublattice $B_2$ with
\begin{equation}
\Psi^s_{+B_2}(y,k)=(\sqrt\pi l)^{-1/2}e^{-(y-y_k)^2/2l^2},
\label{wavefun_n=0}
\end{equation}
and is not perturbed by the left edge ($y=0$) of the ribbon. The only effect of this edge on the wave function is that it becomes zero outside the ribbon and the normalization constant in Eq.~(\ref{wavefun_n=0}) changes accordingly. When the momentum becomes negative and the guiding center $y_k$ moves farther away from the ribbon, the bulk Landau level $n=0$ evolves into a dispersionless branch of strongly localized near the left edge states residing on a single sublattice and layer $B_2$ with the same energy $E=-\Delta$ and described by the wave function
\begin{equation}
\Psi^s_{+B_2}(y,k)=Ce^{-|k|y-y^2/2l^2}\simeq Ce^{-|k|y},
\label{wavefun_dispersionless_0}
\end{equation}
where $C$ is a normalization constant. On a half plane, these states correspond to the exact solution $E=-\Delta$ of dispersion equation~(\ref{hp_zigzag_disp_eq_plus}).

The wave function~(\ref{sol_inf_plane}) of the $n=1$ bulk Landau level has three nonzero components,
\begin{equation}
\Psi^s_+(y,k)
=Ce^{-(y-y_k)^2/2l^2}
\begin{pmatrix}
0 \\
\frac{\epsilon_0^2-(E_1^{-,+}+\Delta)^2}{\sqrt2\epsilon_0\gamma_1} \\
-\frac{E_1^{-,+}+\Delta}{\sqrt2\epsilon_0} \\
(y-y_k)/l
\end{pmatrix}
.
\label{wavefun_n=1}
\end{equation}
This level disperses upwards when the guiding center $y_k$ approaches the left edge, and its energy grows gradually from $E_1^{-,+}$ to $\Delta$. This behavior has been qualitatively described in Ref.~\onlinecite{Mazo2011PRB} by the variational method, using the ``bulklike'' anzatz $\Psi^s_{+A_1}(y,k)\equiv0$. Here, by using the properties of the parabolic cylinder functions, we find the exact asymptotic behavior of this branch at large positive and negative momenta (see Appendix~\ref{Appendix:asymptotes}). For $kl\gg1$, the deviation from the bulk energy $E_1^{-,+}$ is exponentially small and given by Eq.~(\ref{E1asympt}), whereas at $-kl\gg1$ the energy has the following asymptotic behavior:
\begin{equation}
E\simeq\Delta\biggl(1-\frac{\gamma_1^2}{2\hbar^2v_F^2k^2}\biggr),
\label{energy_subgap_0}
\end{equation}
and the corresponding wave function of the resulting edge mode is approximately given by
\begin{equation}
\Psi^s_+(y,k)\simeq Ce^{-|k|y}
\begin{pmatrix}
-\frac{\Delta\gamma_1y}{\hbar^2v_F^2k} \\
\frac{2\hbar v_Fk}{\gamma_1} \\
\frac{2\Delta y}{\hbar v_F} \\
1+2ky
\end{pmatrix}
.
\label{wavefun_subgap_0}
\end{equation}
In contrast to the \emph{single-layer} left-edge dispersionless mode~(\ref{wavefun_dispersionless_0}), this solution has nonzero components corresponding to both graphene layers, and thus will be referred to as a \emph{double-layer} left-edge mode.

For a ribbon of a finite width $W$, there are two more low-energy solutions of the dispersion equations, which are absent in the case of a half plane and correspond to the modes localized near the right edge ($y=W$) of the ribbon. One of them forms a horizontal plateau with energy $E=\Delta$, which does not correspond to any of the bulk Landau levels. This mode is described by the wave function
\begin{equation}
\Psi^s_+(y,k)\simeq Ce^{-\frac{\eta^2}2}
\begin{pmatrix}
\frac{\sqrt{2\pi}\Delta\gamma_1}{\epsilon_0^2}e^{\eta^2}\erfc(-\eta) \\
\frac{\epsilon_0^2-4\Delta^2}{\gamma_1\epsilon_0} \\
-\frac{2\Delta}{\epsilon_0} \\
\sqrt2\eta
\label{wavefun_dispersionless_W_full}
\end{pmatrix}
,
\end{equation}
where $\eta=y/l-kl$ and
\begin{equation}
\erfc(x)=1-\frac2{\sqrt\pi}\int_0^xdt\,e^{t^2}
\end{equation}
is the complementary error function. The boundary condition $\Psi^s_{+B_{1,2}}(W,k)=0$ does not perturb this state noticeably because all its components, except $\Psi^s_{+A_1}$, are localized in the bulk near $y=y_k$. At $kl-W/l\ll-1$ these bulklike components are negligibly small compared to $\Psi^s_{+A_1}(y,k)$, and the normalized wave function~(\ref{wavefun_dispersionless_W_full}) is approximately given by
\begin{equation}
\Psi^s_+(y,k)\simeq \sqrt{2|k'|}e^{-|k'|(W-y)}
\begin{pmatrix}
1 \\
0 \\
0 \\
0
\end{pmatrix}
,\quad k'\equiv k-\frac W{l^2}.
\label{wavefun_dispersionless_W}
\end{equation}
As one can see, this \emph{single-layer} state is localized near the right edge even when the guiding center $y_k$ is deep in the bulk; hence the position wave-vector duality is not applicable in this case. This purely edge state branch is completely analogous to the one that exists in gapped monolayer graphene.~\cite{Gusynin2009PRB}

The energy of another right-edge mode changes from $+\Delta$ to $-\Delta$ as $y_k$ moves into the bulk. At $-k'l\gg1$ it is described by the asymptote (see Appendix~\ref{Appendix:asymptotes})
\begin{equation}
E\simeq\Delta\biggl(1-\frac{\gamma_1^2}{2\hbar^2v_F^2k'^2}\biggr),
\label{energy_subgap_W}
\end{equation}
and the corresponding wave function is approximately given by
\begin{equation}
\Psi^s_+(y,k)\simeq Ce^{-|k'|(W-y)}
\begin{pmatrix}
1+2k'(W-y) \\
-\frac{2\Delta(W-y)}{\hbar v_F} \\
\frac{2\hbar v_Fk'}{\gamma_1} \\
\frac{\Delta\gamma_1(W-y)}{\hbar^2v_F^2k'}
\end{pmatrix}
.
\label{wavefun_subgap_W}
\end{equation}
The position wave-vector duality is not applicable for this \emph{double-layer} right-edge mode mode as well.

In unbiased ($\Delta=0$) bilayer graphene, the spectrum is electron-hole symmetric, with the positive and negative energy solutions related by
\begin{equation}
\Psi^s_\xi(y,k,E)=C
\begin{pmatrix}
\sigma_3 & 0 \\
0 & \sigma_3
\end{pmatrix}
\Psi^s_\xi(y,k,-E).
\end{equation}
The orthogonality of those states implies that the probabilities of finding the electron on each sublattice are equal,
\begin{align}
\int_0^Wdy\,\Bigl(&|\Psi^s_{\xi A_1}|^2+|\Psi^s_{\xi A_2}|^2\Bigr) \nonumber \\
&=\int_0^Wdy\,\Bigl(|\Psi^s_{\xi B_1}|^2+|\Psi^s_{\xi B_2}|^2\Bigr)=\frac12.
\end{align}
At $\Delta=0$ in the $K_+$ valley, the bulk states $n=0$ and $n=1$ with zero energy reside solely on the $B$ sublattice, while the edge states~(\ref{wavefun_dispersionless_W}) and~(\ref{wavefun_subgap_W}) reside on the $A$ sublattice. Therefore, in the range of momenta where the bulk $n=0,1$ solutions are present, these bulk states are hybridized with the right-edge states, so that the probabilities to find the electron in the bulk and at the right edge are equal. Similar mixing of the bulk and edge states occurs in gapless monolayer graphene with zigzag edges.\cite{Brey2006PRB,Abanin2007SSC,Romanovsky2011PRB} In Appendix~\ref{Appendix:asymptotes} we show that when $y_k$ is deep in the bulk ($-k'l\gg1$), the bulk $n=0$ ($n=1$) states admix mainly with the single (double) layer right edge states and also find the dispersion of these mixed bulk-edge modes.

At finite $\Delta$, the spectrum of the lowest energy branches for $y_k$ located near the right edge of the ribbon is characterized by a transition from the distinct bulk and edge branches at $-k'l\gg1$ to the mixed bulk-edge modes on the energy scales $\Delta\ll|E|\ll\epsilon_0$. At $-k'l\gg1$, the deviations of the $n=0$ and $n=1$ level energies from their bulk values are exponentially small; see Eqs.~(\ref{E0_disp_right}) and~(\ref{E1_disp_right}). When $y_k$ moves towards the right edge and these deviations become comparable with the separations between the two levels, the bulk modes start admixing with the corresponding edge modes, so that at $|E|\gtrsim\Delta$ the modes are almost completely hybridized and their energies quickly approach their $\Delta=0$ counterparts (dashed lines in Fig.~\ref{LLLs}). As the guiding center $y_k$ of the bulk states crosses the right edge of the ribbon, these mixed bulk-edge modes evolve further into conventional quantum Hall edge states similar to those at the higher Landau levels.

The bulk wave function~(\ref{wavefun_n=1}) of the level $n=1$ is extended through a larger $y$ interval and therefore is perturbed stronger by the edge than the wave function~(\ref{wavefun_n=0}) of the level $n=0$. Because of this, the level $n=1$ starts dispersing downwards at smaller $k$, and the avoided crossings $\mathcal A1$ and $\mathcal A2$ of the partially mixed bulk $n=1$ and double-layer edge modes with bulk $n=0$ plateau and dispersionless single-layer edge branch are formed (Fig.~\ref{schematic}). The third avoided crossing $\mathcal A3$ occurs when the double-layer right-edge mode intersects either the left-edge double-layer mode [Fig.~\ref{LLLs}(a)] or the bulk $n=1$ plateau [Figs.~\ref{LLLs}(b)--\ref{LLLs}(d)], depending on the magnetic field strength and the width of the ribbon.

In the case $2\Delta\ll\epsilon_0$, the structure of the spectrum at the energy scale $|E|\lesssim\Delta$ is almost independent of the gap parameter. The main effect of decreasing $\Delta$ is that the bulk and edge modes start mixing and approach their $\Delta=0$ counterparts at smaller $k$; see Figs.~\ref{zigzag_fig}(b) and~\ref{zigzag_fig}(c).

\begin{figure}
\includegraphics[width=0.85\columnwidth]{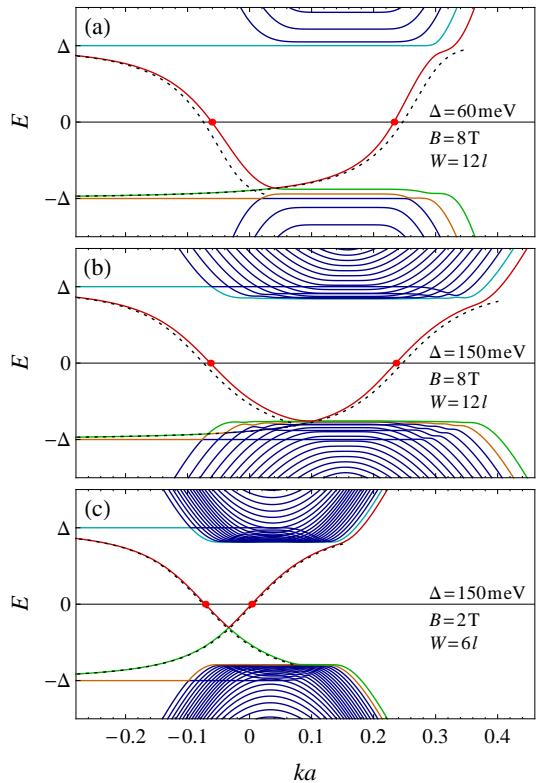}
\caption{Numerical results for the low-energy spectrum in the $K_+$ valley for zigzag ribbon of a constant width $W\simeq110$\,nm at different values of magnetic field and gap parameter $\Delta>\epsilon_0/2$. Dotted lines show the subgap edge modes at $B=0$ (right edge mode is shifted horizontally with $k\to k+W/l^2$ in order to illustrate the effect of a magnetic field). Gapless edge states at $\mu_s=0$ are marked by dots.}
\label{smallB}
\end{figure}

The spacings between higher Landau levels $(|n|\geqslant2)$ decrease with decreasing magnetic field or increasing gate voltage. In particular, when the parameter $\Delta/\epsilon_0$ is increased above the threshold value of $1/2$, the crossings of different Landau levels occur\cite{Pereira2007PRB} and the levels $n=0$ and $n=1$ are no longer the lowest ones. The numerical results for the energy spectrum in this regime are shown in Fig.~\ref{smallB}.

As one can see from Eqs.~(\ref{wavefun_dispersionless_0})--(\ref{wavefun_subgap_0}) and (\ref{wavefun_dispersionless_W})--(\ref{wavefun_subgap_W}), all four edge modes at large momenta do not depend on magnetic field strength. Indeed, the dispersionless single-layer edge modes are exactly given by Eqs.~(\ref{wavefun_dispersionless_0}) and (\ref{wavefun_dispersionless_W}) at $k<0$ in the limit of $B\to0$ ($l\to\infty$). In Appendix~\ref{Appendix:zeroB} we also show that the double-layer edge modes in this limit turn into the subgap edge modes described by the dispersion equation\cite{Li2010PRB}
\begin{equation}
\hbar^2v_F^2(k+\kappa_+)(k+\kappa_-)=(E\mp\Delta)^2,
\label{disp_eqn_zeroB}
\end{equation}
where
\begin{align}
&\kappa_\pm^2 = k^2-\frac{E^2+\Delta^2\pm i\sqrt{\gamma_1^2(\Delta^2-E^2)-4E^2\Delta^2}}{\hbar^2v_F^2}, \nonumber \\ 
&\re\kappa_\pm>0,
\label{kappa}
\end{align}
and the upper (lower) sign in Eq.~(\ref{disp_eqn_zeroB}) is chosen for the left (right) edge mode.

\begin{figure}
\includegraphics[width=0.9\columnwidth]{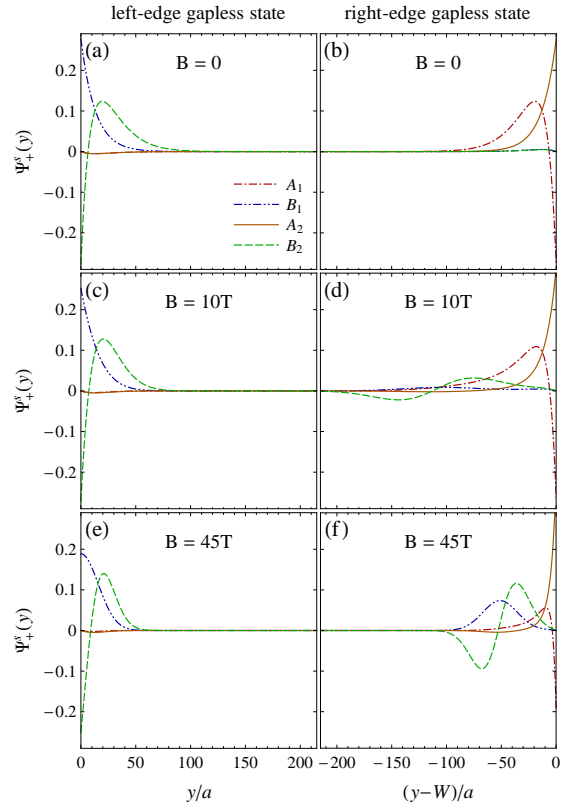}
\caption{Wave functions of zero-energy states in a wide zigzag ribbon in the $K_+$ valley at $\Delta=10$\,meV and different values of a magnetic field.}
\label{wavefun}
\end{figure}

Moderate magnetic fields change the dispersion of the left-edge states determined by Eq.~(\ref{disp_eqn_zeroB}) only slightly, whereas the right-edge mode becomes partially hybridized with the states of the bulk $n=1$ Landau level (Fig.~\ref{schematic}) and shifted horizontally with $\delta k=W/l^2$.

At large negative momenta all four branches continue as their counterparts in the $K_-$ valley, with the energy spectrum and the corresponding wave functions related by Eqs.~(\ref{zigzag_E_symmetry}) and~(\ref{zigzag_psi_symmetry}). Thus the width of each edge state mode is equal to $2\pi/(3a)$, i.e., the spacing between $K_+$ and $K_-$ points (although not captured by the continuum model, this fact can be directly seen from the tight-binding calculations\cite{Castro2007PRL,Mazo2011PRB}). The finiteness of the branches can be taken into account within a continuum model by introducing the momentum cutoffs $k^{(\prime)}_c=-\pi/(3a)$ for the left (right) edge modes in the $K_+$ valley, and the corresponding cutoffs in the $K_-$ valley. These cutoffs can be seen in Figs.~\ref{zigzag_fig}(d) and~\ref{zigzag_fig}(e).

The electronic structure described above implies that current-carrying gapless edge states (states located at the Fermi level, $E=\mu_s\equiv\mu+s\Delta_Z$, $s=\pm$, where $\mu$ is the chemical potential and $\Delta_Z=\mu_BB\approx0.06\,B[{\rm T}]$\,meV is the Zeeman energy) are always present for all realistic magnetic fields. At $\mu=0$ and $\Delta_Z\ll\Delta$, there are two gapless states in each valley, which carry currents in opposite directions on a given edge.\cite{Castro2007PRL,Mazo2011PRB} Zero-energy states exist if the energy of the $n=1$ Landau level $E_1^{-,+}$ is negative. This condition is violated only at ultrahigh magnetic fields [see Fig.~\ref{LLLs}(e)] exceeding $B_{\rm cr}$, the exact value of which can be determined from Eq.~(\ref{bulk_LL_eq_n1}) at $E=0$,
\begin{equation}
\epsilon_0^2=\gamma_1^2+\Delta^2,
\end{equation}
which implies
\begin{equation}
B_{\rm cr}\,[\mathrm T]\approx123\bigl[1+(2.5\Delta[\mathrm{eV}])^2\bigr].
\label{Bcr}
\end{equation}

In Fig.~\ref{wavefun}, the evolution of the gapless states at $\mu_s=0$ with increasing magnetic field is shown: the left-edge state in the $K_+$ valley remains almost unchanged, while the right-edge state becomes partially admixed with the $n=1$ bulk state.

\section{Armchair edges}
\label{secIV}

\subsection{General solution with \textit{y} translational invariance}

In the armchair edge case, the solution~(\ref{gen_sol}) has to be modified as follows. We consider the edge(s) along the $y$ axis and choose the gauge $\mathbf A=(0,Bx)$. The wave functions are plane waves in the $y$ direction,
\begin{equation}
\Psi^s_\xi(x,y)=e^{iky}\widetilde\Psi^s_\xi(x,k),
\end{equation}
where the envelope functions $\widetilde\Psi^s_\xi(x,k)\equiv\widetilde\Psi^s_\xi(\eta)$ depend only on a single combination of the variables, $\eta=x/l+kl$, and satisfy the equation
\begin{equation}
\xi
\begin{pmatrix}
\Delta & -i\epsilon_0\hat a & 0 & 0 \\
i\epsilon_0\hat a^\dag & \Delta & \xi\gamma_1 & 0 \\
0 & \xi\gamma_1 & -\Delta & -i\epsilon_0\hat a \\
0 & 0 & i\epsilon_0\hat a^\dag & -\Delta
\end{pmatrix}
\widetilde\Psi^s_\xi(\eta)=E\widetilde\Psi^s_\xi(\eta),
\end{equation}
or, equivalently, Eq.~(\ref{E_eig_eq}) with the solution related by the unitary transformation
\begin{equation}
\Psi^s_\xi(\eta)=
\hat S^{-1}\widetilde\Psi^s_\xi(\eta), \qquad
\hat S=
\begin{pmatrix}
i & 0 & 0 & 0 \\
0 & 1 & 0 & 0 \\
0 & 0 & 1 & 0 \\
0 & 0 & 0 & -i
\end{pmatrix}
.
\label{psi_transform}
\end{equation}
Therefore, the general solution in this case is
\begin{equation}
\widetilde\Psi^s_\xi(\eta)=\sum_{i=1}^4C_i^\xi\widetilde\Phi_\xi^{(i)}(\eta),
\end{equation}
where
\begin{equation}
\widetilde\Phi_\xi^{(i)}(\eta)=\hat S\Phi_\xi^{(i)}(\eta).
\end{equation}

\subsection{Dispersion equation for half plane}

On a semi-infinite plane $x>0$, the normalizable wave functions are given in terms of only the $U(a,z)$ function, and the solution is
\begin{equation}
\widetilde\Psi^s_\xi(\eta)=C_1^\xi\widetilde\Phi_\xi^{(1)}(\eta)+C_2^\xi\widetilde\Phi_\xi^{(2)}(\eta).
\label{gen_sol_hp_armchair}
\end{equation}

At the armchair edge $x=0$ (which corresponds to $\eta=kl$) of a semi-infinite plane, the appropriate boundary conditions for the continuum model are that the wave function should vanish on both sublattices~\cite{Brey2006PRB,Abanin2006PRL} and layers:
\begin{equation}
\sum_{\xi=\pm}\widetilde\Psi^s_{\xi A_i}(kl)=\sum_{\xi=\pm}\widetilde\Psi^s_{\xi B_i}(kl)=0,\qquad i=1,2.
\label{armchair_half_plane_BC}
\end{equation}
These boundary conditions mix the components of the wave function from two valleys $\xi=\pm$, and we have the system of four equations
\begin{equation}
\begin{array}{l}
\sum_{\xi=\pm}\bigl[C_1^\xi\widetilde\Phi_{\xi A_i}^{(1)}(kl)
+C_2^\xi\widetilde\Phi_{\xi A_i}^{(2)}(kl)\bigr]=0, \smallskip \\
\sum_{\xi=\pm}\bigl[C_1^\xi\widetilde\Phi_{\xi B_i}^{(1)}(kl)
+C_2^\xi\widetilde\Phi_{\xi B_i}^{(2)}(kl)\bigr]=0, \quad i=1,2,
\end{array}
\end{equation}
which has a nontrivial solution when the corresponding determinants of coefficient functions are zero. This condition can be written as
\begin{equation}
\frac1{(\lambda_2^+-\lambda_1^+)(\lambda_2^--\lambda_1^-)}
\begin{vmatrix}
a_1^+ & a_2^+ & d_1^- & d_2^- \\
b_1^+ & b_2^+ & c_1^- & c_2^- \\
c_1^+ & c_2^+ & b_1^- & b_2^- \\
d_1^+ & d_2^+ & a_1^- & a_2^- 
\end{vmatrix}
=0,
\label{hp_armchair_disp_eq}
\end{equation}
where
\begin{align}
a_i^\pm &= \nu_\pm(\lambda^\pm_i)U\biggl(\frac32-\lambda^\pm_i,\sqrt2kl\biggr), \nonumber \\
b_i^\pm &= \frac{E\pm\Delta}{\epsilon_0}U\biggl(\frac12-\lambda^\pm_i,\sqrt2kl\biggr), \nonumber \\
c_i^\pm &= \frac{(E\pm\Delta)^2-\epsilon_0^2\lambda^\pm_i}{\epsilon_0\gamma_1}
U\biggl(\frac12-\lambda^\pm_i,\sqrt2kl\biggr), \nonumber \\
d_i^\pm &= -U\biggl(-\frac12-\lambda^\pm_i,\sqrt2kl\biggr).
\end{align}
It is easy to see that the left-hand side of the dispersion equation~(\ref{hp_armchair_disp_eq}) does not depend on the sign of the energy; therefore, the spectrum is symmetric under the transformation $E\to-E$. The corresponding wave functions are transformed as
\begin{equation}
\Psi^s_\xi(x,k)
\to
\xi C
\begin{pmatrix}
\sigma_0 & 0 \\
0 & -\sigma_0
\end{pmatrix}
\Psi^s_{-\xi}(x,k),
\label{esignchange}
\end{equation}
where $\sigma_0$ is a unit $2\times2$ matrix.

\subsection{Dispersion equations for ribbon}

\begin{figure*}
\includegraphics[width=\textwidth]{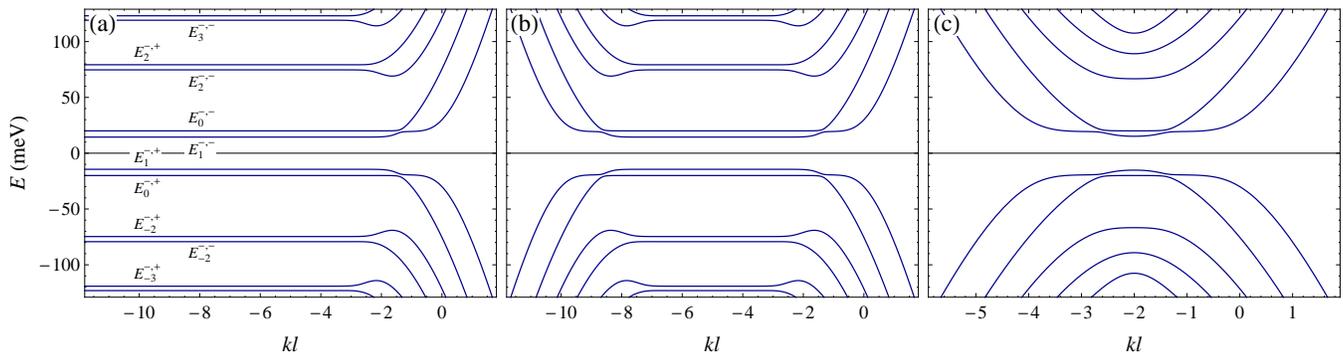}
\caption{Numerical results for the low-energy spectrum in bilayer graphene with armchair edge(s) at $B=20\,\mathrm T$ and $\Delta=20\,\mathrm{meV}$: (a)~half plane, (b)~ribbon of the width $W=10l$, (c)~ribbon of the width $W=4l$. On panel (a), the bulk Landau level energies $E_n^{\kappa,\xi}$ are indicated.}
\label{armchair_fig}
\end{figure*}

The boundary conditions at two armchair edges $x=0$ and $x=W$ (corresponding to $\eta=kl$ and
$\eta=W/l+kl$, respectively) are
\begin{align}
\sum_{\xi=\pm}&\widetilde\Psi^s_{\xi A_i}(kl)
=\sum_{\xi=\pm}\widetilde\Psi^s_{\xi B_i}(kl) 
=\sum_{\xi=\pm}\widetilde\Psi^s_{\xi A_i}(W/l+kl) \nonumber \\
&=\sum_{\xi=\pm}\widetilde\Psi^s_{\xi B_i}(W/l+kl)=0,\qquad i=1,2;
\end{align}
i.e., the boundary condition~(\ref{armchair_half_plane_BC}) is imposed at both ribbon edges. These valley-mixing boundary conditions lead to the system of eight equations
\begin{equation}
\begin{array}{l}
\sum_{j=1}^4\sum_{\xi=\pm}C_j^\xi\widetilde\Phi_{\xi A_i}^{(j)}(kl)=0, \medskip \\
\sum_{j=1}^4\sum_{\xi=\pm}C_j^\xi\widetilde\Phi_{\xi B_i}^{(j)}(kl)=0, \medskip \\
\sum_{j=1}^4\sum_{\xi=\pm}C_j^\xi\widetilde\Phi_{\xi A_i}^{(j)}(W/l+kl)=0, \medskip \\
\sum_{j=1}^4\sum_{\xi=\pm}C_j^\xi\widetilde\Phi_{\xi B_i}^{(j)}(W/l+kl)=0, \qquad i=1,2.
\end{array}
\end{equation}
Equating the determinant of the above system to zero, one gets the dispersion equation for the ribbon with armchair edges. After some algebra, it can be written as
\begin{equation}
\frac1{(\lambda_2^+-\lambda_1^+)^2(\lambda_2^--\lambda_1^-)^2}
\begin{vmatrix}
Z_1(kl) & Z_1(W/l+kl) \\
Z_2(kl) & Z_2(W/l+kl)
\end{vmatrix}
=0,
\label{r_armchair_disp_eq}
\end{equation}
where
\begin{equation}
Z_i(\eta)=
\begin{pmatrix}
X_i^+(\eta) & Y_i^+(\eta) \\
Y_i^-(\eta) & X_i^-(\eta)
\end{pmatrix}
\end{equation}
is a $4\times4$ matrix constructed with $2\times2$ blocks $X_i^\pm(\eta)$, $Y_i^\pm(\eta)$
defined in Eqs.~(\ref{X}) and~(\ref{Y}).

The spectrum is symmetric both with respect to the change of the sign of energy, with the wave function being transformed according to Eq.~(\ref{esignchange}), and the transformation $k\to-W/l^2-k$, with the wave functions transforming as
\begin{equation}
\Psi^s_\xi(x,k)
\to
C
\begin{pmatrix}
\sigma_3 & 0 \\
0 & -\sigma_3
\end{pmatrix}
\Psi^s_\xi(W-x,-W/l^2-k).
\end{equation}

\subsection{Numerical results for the spectra}

We numerically solve the dispersion equations~(\ref{hp_armchair_disp_eq}) and~(\ref{r_armchair_disp_eq}) in the case of the semi-infinite plane and the finite-width ribbon, respectively. The examples of the spectra showing a few lower Landau levels in the case of a half plane and two different widths of the ribbon are shown in Fig.~\ref{armchair_fig}.

On a half plane, the branches of the spectrum asymptotically approach the bulk Landau levels given by Eqs.~(\ref{bulk_LL_eq})--(\ref{bulk_LL_n0}). The states corresponding to these bulk asymptotes (plateaus) are localized in the bulk and centered along the $x$ direction at $x_k=-kl^2$ [the position wave-vector duality in the gauge $A=(0,Bx)$]. These states are bulk states predominantly concentrated on a single valley $K_\xi$ and approximately described by the infinite-plane solutions~(\ref{sol_inf_plane}) with $\eta=x/l+kl$, transformed according to Eq.~(\ref{psi_transform}). The same is true for wide ribbons $W\gg l$ [see Fig.~\ref{armchair_fig}(b)], where the plateaus closely approaching the bulk Landau levels are developed. For a given branch, all states to the left (right) of the bulk plateau are localized in the vicinity of the edge $x=W$ ($x=0$).

Similarly to the zigzag edge case, the widths of the Landau level plateaus with larger $|n|$ are smaller due to the growing localization lengths of the bulk states. For the same reason, the avoided crossings of the branches corresponding to $n=0$ and $n=1$ Landau levels are formed.

\begin{figure}[b]
\includegraphics[width=0.75\columnwidth]{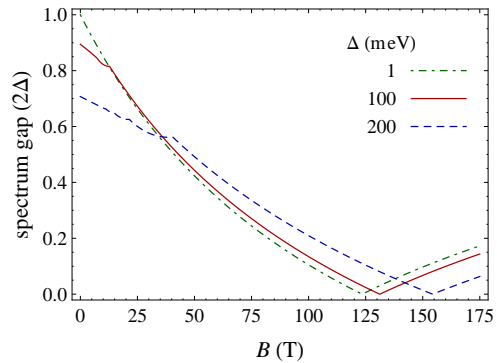}
\caption{Gap in the spectrum (in the units of $2\Delta$) in the armchair edge case
as a function of a magnetic field at different values of $\Delta$.}
\label{E1fig}
\end{figure}

In contrast to the zigzag case, there are no additional edge state branches, and gapless quantum Hall edge states exist only when the chemical potential (including the Zeeman energy) $\mu_s$ exceeds the spectrum gap. The latter is determined by the energy of the lowest-lying Landau level. At $2\Delta<\epsilon_0$ the lowest level is $n=1$ and the spectrum gap is equal to $2|E_1^{-,+}|$, whereas at $2\Delta>\epsilon_0$, the gap in the spectrum is determined by the higher Landau levels. This gap decreases monotonically with increasing $B$, closes at $B=B_{\rm cr}$ given by Eq.~(\ref{Bcr}) and then grows again (Fig.~\ref{E1fig}). For all experimentally accessible values of magnetic field and layer asymmetry gap, the size of the spectrum gap varies between $\Delta$ and $2\Delta$.

\section{Conclusions}
\label{secV}

In summary, we studied the spectrum of biased bilayer graphene with zigzag or armchair edges in a magnetic field within the continuum four-band model. We derived the general analytic solution for the wave functions in a ribbon with zigzag or armchair edges. For both edge types, the exact dispersion equations were written in terms of the parabolic cylinder functions. Solving these dispersion equations numerically, we obtained the spectra of noninteracting electrons in a bilayer graphene ribbon or semi-infinite plane at different values of magnetic field and layer asymmetry gap induced by the gate voltage.

The edge state spectrum close to the charge neutrality point is found to depend strongly on the edge type. Zigzag edges are shown to support zero-energy edge states propagating in opposite directions in the two valleys, in agreement with the previous tight-binding studies.\cite{Castro2007PRL,Mazo2011PRB,Wu2012PRB,Zhang2012PRB} Some of these states remain almost unchanged when the magnetic field is turned on and increased up to the highest values currently accessible in experiments, whereas the others become partially hybridized with the bulk state of the $n=1$ Landau level. The behavior of the lowest-energy spectrum branches at large momenta as well as their zero magnetic field limit has been investigated in detail by using the asymptotic properties of the parabolic cylinder functions. In contrast, the spectrum of armchair bilayer graphene ribbon is gapped and zero-energy edge states are absent. The gap in the edge state spectrum is equal to the gate-voltage-induced bulk gap (Fig.~\ref{E1fig}), the size of which is determined both by the gap parameter $\Delta$ and the magnetic field strength.

The obtained structure of edge states suggests the following implications on transport properties of a bilayer graphene ribbon in the quantum Hall regime at zero filling. When the spin splitting is less than the gate-induced gap, the current-carrying gapless edge states are present only in the case of zigzag edges. In clean samples with ideal zigzag edges, these states should form the conducting channels resulting in the metallic state with the finite two-terminal or four-terminal longitudinal conductance equal to $4e^2/h$ (corresponding to the states with different spin projections at each edge of the ribbon). However, these edge channels are not protected against the backscattering in the presence of the valley-mixing edge disorder, therefore their contribution to the conductance can be sensitive to the edge structure of real samples.\cite{Li2011NP} When the spin splitting exceeds the gate-induced gap, metallic behavior is expected regardless of the edge type due to the counterpropagating gapless edge states with opposite spin projections.\cite{Abanin2006PRL,Fertig2006PRL,Abanin2007SSC,Kharitonov2012}

While the present paper deals with the case of nonzero layer asymmetry gap and the spin splitting, it would be interesting to extend our analysis to a more general set of order parameters, similarly to the studies of Refs.~\onlinecite{Gusynin2008PRB} and~\onlinecite{Gusynin2009PRB} in monolayer graphene. Experiments with bilayer graphene in magnetic fields at the charge neutrality point reveal different phases with spontaneously broken symmetries,\cite{Weitz2010S,Velasco2012NN,Maher2013NP} and the knowledge of low-energy edge state structure is essential for identifying the true nature of the ground state in each phase.\cite{Maher2013NP,Kharitonov2012}

\section{Acknowledgments}
The author is sincerely grateful to V.~A.~Miransky for the formulation of the problem and useful discussions. The work was supported by the Natural Sciences and Engineering Research Council of Canada and by the Ontario Graduate Scholarship program.

\appendix

\section{Derivation of the general solution}
\label{Appendix:solution}

Let us start with Eq.~(\ref{E_eig_eq}) for the $K_+$ valley, which is written in components as
\begin{align}
\epsilon_0\hat a\Psi_{+B_1}^s &= -(E-\Delta)\Psi_{+A_1}^s, \nonumber \\
\epsilon_0\hat a^\dag\Psi_{+A_1}^s &= -(E-\Delta)\Psi_{+B_1}^s+\gamma_1\Psi_{+A_2}^s, \nonumber \\
\epsilon_0\hat a\Psi_{+B_2}^s &= -(E+\Delta)\Psi_{+A_2}^s+\gamma_1\Psi_{+B_1}^s, \nonumber \\
\epsilon_0\hat a^\dag\Psi_{+A_2}^s &= -(E+\Delta)\Psi_{+B_2}^s. \label{eqs_K+}
\end{align}
Eliminating $\Psi_{+A_1}^s$, $\Psi_{+A_2}^s$, and $\Psi_{+B_2}^s$ leads to the fourth-order differential equation
\begin{align}
\bigl[\epsilon_0^2\hat a\hat a^\dag-(E+\Delta)^2\bigr]
\bigl[\epsilon_0^2\hat a^\dag&\hat a-(E-\Delta)^2\bigr]\Psi_{+B_1}^s \nonumber \\
&=\gamma_1^2(E^2-\Delta^2)\Psi_{+B_1}^s,
\end{align}
which admits the factorization~\cite{Pereira2007PRB}
\begin{equation}
\bigl(\partial_\eta^2-\eta^2-1+2\lambda_1^+\bigr)
\bigl(\partial_\eta^2-\eta^2-1+2\lambda_2^+\bigr)
\Psi_{+B_1}^s(\eta)=0
\label{diff_eq_factorized}
\end{equation}
with $\lambda_{1,2}^\xi$ defined in Eq.~(\ref{lambdas}).
Therefore, the general solutions of equations
\begin{equation}
\bigl(\partial_\eta^2-\eta^2-1+2\lambda_i^+\bigr)
\Psi_{+B_1}(\eta)=0,
\qquad
i=1,2,
\end{equation}
given by the pairs of the linearly independent parabolic cylinder functions $U(1/2-\lambda_i,\sqrt2\eta)$, $V(1/2-\lambda_i,\sqrt2\eta)$, also satisfy Eq.~(\ref{diff_eq_factorized}). Combining these solutions gives
\begin{align}
&\Psi_{+B_1}(\eta)=C_1U\Bigl(\frac12-\lambda_1^+,\sqrt2\eta\Bigr)+
C_2U\Bigl(\frac12-\lambda_2^+,\sqrt2\eta\Bigr) \nonumber \\
&\quad+C_3V\Bigl(\frac12-\lambda_1^+,\sqrt2\eta\Bigr)+
C_4V\Bigl(\frac12-\lambda_2^+,\sqrt2\eta\Bigr).
\label{sol_B1}
\end{align}
All four functions in the above equation are linearly independent at $\lambda_1^+\ne\lambda_2^+$, which can be proved in the following way. The functions $U(a_1,z)$, $U(a_2,z)$, $V(a_1,z)$, $V(a_2,z)$, where $a_{1,2}=-1/2-\lambda^+_{1,2}$ and $z=\sqrt2\eta$, are the solutions of differential equation~(\ref{diff_eq_factorized}) which has zero coefficient at the third derivative term. This implies\cite{ODE_book} that the Wronskian of these functions is equal to some constant, dependent on the parameters $a_{1,2}$. The value of this constant can be found, for example, by evaluating the Wronskian at $z\to+\infty$ and using the asymptotic expressions for the parabolic cylinder functions\cite{Abramowitz}
\begin{align}
U\biggl(a-\frac12,z\biggr) &= e^{-\frac{z^2}4}z^{-a}\biggl[1-\frac{a(a+1)}{2z^2} \nonumber \\
&\quad+\frac{a(a+1)(a+2)(a+3)}{8z^4}-\dots\biggr], \label{DasymptPositive} \\
V\biggl(a+\frac12,z\biggr) &= \sqrt{\frac2\pi}e^{\frac{z^2}4}z^a\biggl[1+\frac{a(a-1)}{2z^2} \nonumber \\
&\quad+\frac{a(a-1)(a-2)(a-3)}{8z^4}+\dots\biggr]. \label{DasymptNegative}
\end{align}
The result is
\begin{equation}
\mathcal W\bigl[U(a_1,z),U(a_2,z),V(a_1,z),V(a_2,z)\bigr]=\frac2\pi(a_1-a_2)^2.
\label{Wronskian}
\end{equation}
At $\lambda_1^+=\lambda_2^+$ [this equality is possible in the case $\epsilon_0^4<4\Delta^2(\gamma_1^2+4\Delta^2)$], the solutions in Eq.~(\ref{sol_B1}) are not linearly independent. This can be fixed by rearranging terms as
\begin{align}
&\Psi_{+B_1}(\eta)=\widetilde C_1U\bigl(1/2-\lambda_1^+,\sqrt2\eta\bigr) \nonumber \\
&\;+\widetilde C_2\frac{U\bigl(1/2-\lambda_1^+,\sqrt2\eta\bigr)-U\bigl(1/2-\lambda_2^+,\sqrt2\eta\bigr)}
{\lambda_1^+-\lambda_2^+} \nonumber \\
&\;+\widetilde C_3V\bigl(1/2-\lambda_1^+,\sqrt2\eta\bigr) \nonumber \\
&\;+\widetilde C_4\frac{V\bigl(1/2-\lambda_1^+,\sqrt2\eta\bigr)-V\bigl(1/2-\lambda_2^+,\sqrt2\eta\bigr)}
{\lambda_1^+-\lambda_2^+},
\end{align}
where the resulting four solutions are linearly independent at arbitrary $\lambda_{1,2}^+$, with the corresponding Wronskian
\begin{align}
\mathcal W\biggl[U(a_1,z), \frac{U(a_2,z)-U(a_1,z)}{a_2-a_1}, V(a_1,z),& \nonumber \\
\frac{V(a_2,z)-V(a_1,z)}{a_2-a_1}\biggr]&=\frac2\pi.
\end{align}

The remaining components of $\Psi_+^s$ can be obtained from Eqs.~(\ref{eqs_K+}) and~(\ref{sol_B1}) by using the recurrence relations for the parabolic cylinder functions, which are written in terms of operators $\hat a$ and $\hat a^\dag$ as
\begin{align}
\hat aU(-\lambda-1/2,\sqrt2\eta)&=\lambda\,U(-\lambda+1/2,\sqrt2\eta), \label{recurrU} \\
\hat a^\dag U(-\lambda+1/2,\sqrt2\eta)&=U(-\lambda-1/2,\sqrt2\eta),  \\
\hat aV(-\lambda-1/2,\sqrt2\eta)&=V(-\lambda+1/2,\sqrt2\eta), \\
\hat a^\dag V(-\lambda+1/2,\sqrt2\eta)&=\lambda\,V(-\lambda-1/2,\sqrt2\eta). 
\end{align}
The overall factors for the solutions are chosen in such a way that no singularities arise at $E=\pm\Delta$. This leads to expressions~(\ref{gen_sol})--(\ref{h}) with $\xi=+$. The corresponding solutions for the $K_-$ valley can be obtained by making the formal replacement $E\to-E$, $\gamma_1\to-\gamma_1$.

\section{Large momentum asymptotes}
\label{Appendix:asymptotes}

Here we consider the case of a wide zigzag ribbon ($W\gg l$) so that the influence of the right edge can be neglected and derive the large momentum asymptotes for the modes in the $K_+$ valley localized near the left edge by using the corresponding dispersion equation~(\ref{hp_zigzag_disp_eq_plus}) for the semi-infinite plane. Assuming $\lambda^\xi_1\ne\lambda^\xi_2$ and $E\ne\pm\Delta$ it can be written as
\begin{equation}
\frac{(\lambda_1^+-1)w_{\lambda_1^+-1}\bigl(\sqrt2kl\bigr)}{(E-\Delta)^2-(\lambda_1^+-1)\epsilon_0^2}=
\frac{(\lambda_2^+-1)w_{\lambda_2^+-1}\bigl(\sqrt2kl\bigr)}{(E-\Delta)^2-(\lambda_2^+-1)\epsilon_0^2},
\label{hp_disp_eq_mod_plus}
\end{equation}
where
\begin{equation}
w_\lambda(z)=\frac{U(1/2-\lambda,-z)}{U(-1/2-\lambda,-z)}.
\end{equation}
For the modes in the $K_+$ valley localized at the right edge of the ribbon, we employ the half-plane dispersion equation~(\ref{hp_zigzag_disp_eq_minus}) for the $K_-$ valley, which can be written as
\begin{align}
\bigl[(E-\Delta)^2&-\lambda_1^-\epsilon_0^2\bigr]w_{\lambda_1^-}\bigl(\sqrt2kl\bigr) \nonumber \\
&=\bigl[(E-\Delta)^2-\lambda_2^-\epsilon_0^2\bigr]w_{\lambda_2^-}\bigl(\sqrt2kl\bigr),
\label{hp_disp_eq_mod_minus}
\end{align}
and use the correspondence~(\ref{zigzag_E_symmetry}) and~(\ref{zigzag_psi_symmetry}) between the solutions in different valleys.

We are interested in the large $kl$ asymptotics of the solutions of Eqs.~(\ref{hp_disp_eq_mod_plus}) and~(\ref{hp_disp_eq_mod_minus}) in the case when at least one of the parameters $\lambda^\xi_{1,2}$ approaches some integer value. This corresponds to the spectrum near the bulk Landau level plateaus or the low-energy edge modes with horizontal asymptotes $E\to\pm\Delta$. Using asymptotic expansions~(\ref{DasymptPositive}) and~(\ref{DasymptNegative}) and relations~(\ref{UDrelation}) and~(\ref{VDrelation}) between different parabolic cylinder functions, we get at $z\gg1$, $|\epsilon|\ll1$:
\begin{align}
w_{n+\epsilon}(z)&\simeq-z\frac{\sqrt{2\pi}(n-1)!\epsilon-z^{2n-1}e^{-z^2/2}}{\sqrt{2\pi}n!\epsilon-z^{2n+1}e^{-z^2/2}},
\nonumber \\
&\hspace{37mm} n=1,2,3,\dots,
\label{w_asympt_1}
\end{align}
\begin{equation}
w_\epsilon(z)\simeq\frac{\sqrt{2\pi}z}{ze^{-z^2/2}-(1+1/z^2)\epsilon\sqrt{2\pi}},
\label{w_asympt_2}
\end{equation}
\begin{equation}
w_\lambda(z)\simeq-\frac z\lambda\biggl(1-\frac{\lambda+1}{z^2}\biggr),
\quad
\lambda\ne0,1,2,\dots,
\label{w_asympt_3}
\end{equation}
\begin{equation}
w_\lambda(-z)\simeq\frac1z\biggl(1+\frac{\lambda-1}{z^2}\biggr).
\label{w_asympt_4}
\end{equation}
Using the above approximations, we find the asymptotic form of the solutions to dispersion equations~(\ref{hp_disp_eq_mod_plus}) and (\ref{hp_disp_eq_mod_minus}) at $|k|l\gg1$.
The corresponding wave functions are then obtained from the half-plane solution satisfying the boundary conditions, which can be written as
\begin{equation}
\Psi^s_\xi(\eta)=C_1^\xi f^\xi_{\lambda_1^\xi}(\eta)+C_2^\xi f^\xi_{\lambda_2^\xi}(\eta)
\label{wavefun_hp_soln}
\end{equation}
with
\begin{align}
C_j^+&=(-1)^j\bigl[U(1/2-\lambda_j^+,-\sqrt2kl)\bigr]^{-1}, \nonumber \\
C_j^-&=(-1)^j\bigl[U(-1/2-\lambda_j^-,-\sqrt2kl)\bigr]^{-1}.
\end{align}

Let us start with dispersion equation~(\ref{hp_disp_eq_mod_plus}) for the $K_+$ valley. For the double-layer edge mode ($E\simeq\Delta$, $\lambda_{1(2)}^+\simeq1$, $\lambda_{2(1)}^+\simeq4\Delta^2/\epsilon_0^2$, $-kl\gg1$) we use Eq.~(\ref{w_asympt_4}) and arrive at
\begin{equation}
E\simeq\Delta\biggl(1-\frac{\gamma_1^2}{\epsilon_0^2k^2l^2}\biggr),
\label{dlasympt}
\end{equation}
which is equivalent to Eq.~(\ref{energy_subgap_0}). Note that both the $n=0$ Landau level and the single-layer edge mode are strictly dispersionless with their energy $E=-\Delta$ being an exact solution of the half-plane dispersion equation~(\ref{hp_zigzag_disp_eq_plus}). For the dispersion near the bulk $n=1$ level [$E\simeq E_1^{-,+}$, $\lambda_{1(2)}^+\simeq1$, $\lambda_{2(1)}^+\simeq2(E_1^2+\Delta^2)/\epsilon_0^2$, $kl\gg1$] we use Eqs.~(\ref{w_asympt_2}) and~(\ref{w_asympt_3}) and arrive at
\begin{align}
E\simeq E_1^{-,+}+\frac{(E_1^{-,+}-\Delta)^2\epsilon_0^2kl\,e^{-k^2l^2}}
{2\sqrt\pi[\Delta\gamma_1^2+(E_1^{-,+}+\Delta)(E_1^{-,+}-\Delta)^2]}.
\label{E1asympt}
\end{align}

In the $K_-$ valley, we consider only the case $kl\gg1$. For the double-layer edge mode ($E\simeq\Delta$, $\lambda_1^-\simeq0$, $\lambda_2^-\simeq1+4\Delta^2/\epsilon_0^2$) we use Eqs.~(\ref{w_asympt_2}) and~(\ref{w_asympt_3}) and arrive at Eq.~(\ref{dlasympt}), which translates into the asymptotic formula~(\ref{energy_subgap_W}) for the right double-layer edge mode in the $K_+$ valley.

For the single-layer edge mode ($E\simeq-\Delta$, $\lambda_{1(2)}^-\simeq1$, $\lambda_{2(1)}^-\simeq4\Delta^2/\epsilon_0^2$) we use Eqs.~(\ref{w_asympt_1}) with $n=1$ and~(\ref{w_asympt_3}), which leads to
\begin{equation}
E\simeq-\Delta\biggl(1+\frac{\epsilon_0^2(\epsilon_0^2-4\Delta^2)kl\,e^{-k^2l^2}}{2\sqrt\pi\Delta^2\gamma_1^2}\biggr).
\label{dispersion_of_dispersionless}
\end{equation}

For the bulk $n=0$ level ($E\simeq\Delta$, $\lambda_1^-\simeq0$, $\lambda_2^-\simeq1+4\Delta^2/\epsilon_0^2$) we use Eqs.~(\ref{w_asympt_2}) and~(\ref{w_asympt_3}) and arrive at
\begin{equation}
E\simeq\Delta\biggl(1+\frac{\epsilon_0^4k^3l^3e^{-k^2l^2}}{\sqrt\pi\Delta^2\gamma_1^2}\biggr).
\label{E0_disp_right}
\end{equation}

For the bulk $n=1$ level [$E\simeq E_1^{-,-}$, $\lambda_{1(2)}^-\simeq1$, $\lambda_{2(1)}^-\simeq2(E_1^2+\Delta^2)/\epsilon_0^2$] we use Eqs.~(\ref{w_asympt_1}) with $n=1$ and~(\ref{w_asympt_3}) and arrive at
\begin{align}
E&\simeq E_1^{-,-}+\frac1{\sqrt\pi}\biggl(\frac{E_1^{-,-}+\Delta}{E_1^{-,-}-\Delta}\biggr)^2 \nonumber \\
&\qquad\quad\times\frac{\epsilon_0^4k^3l^3e^{-k^2l^2}}
{(E_1^{-,-}-\Delta)(E_1^{-,-}+\Delta)^2-\Delta\gamma_1^2}.
\label{E1_disp_right}
\end{align}

The exponentially small deviations from the bulk Landau levels [Eqs.~(\ref{E1asympt}), (\ref{E0_disp_right}), and (\ref{E1_disp_right})] and the dispersionless edge state branch [Eq.~(\ref{dispersion_of_dispersionless})] are accompanied by exponentially small corrections to the corresponding wave functions, while for the edge modes with dispersion~(\ref{dlasympt}), the wave functions in the $K_+$ valley are given by Eqs.~(\ref{wavefun_subgap_0}) and~(\ref{wavefun_subgap_W}).

In the unbiased case ($\Delta=0$), the spectrum is electron-hole symmetrical and the bulk Landau levels $n=0$ and $n=1$ are degenerate. In the vicinity of this degenerate level ($E\simeq0$, $kl\gg1$), one has
\begin{align}
\lambda_1^-&=1+\frac{E^2}{\epsilon_0^2}\biggl(1+\frac{\gamma_1^2}{\epsilon_0^2}\biggr)
+\mathcal O\bigl(E^4/\epsilon_0^4\bigr), \\
\lambda_2^-&=\frac{E^2}{\epsilon_0^2}\biggl(1-\frac{\gamma_1^2}{\epsilon_0^2}\biggr)
+\mathcal O\bigl(E^4/\epsilon_0^4\bigr).
\end{align}
Using Eqs.~(\ref{w_asympt_1}) and~(\ref{w_asympt_2}), we find the two pairs of approximate solutions of the dispersion equation~(\ref{hp_disp_eq_mod_minus}) at $kl\gg1$,
\begin{equation}
E^{(a)}_\pm=\pm\frac{\epsilon_0^2\,e^{-k^2l^2/2}}{\pi^{1/4}\gamma_1\sqrt{2kl}}\Bigl[1+\mathcal O\bigl((kl)^{-2}\bigr)\Bigr],
\label{E_n0_sle}
\end{equation}
and
\begin{equation}
E^{(b)}_\pm=\pm\frac{\epsilon_0\gamma_1}{\pi^{1/4}}\sqrt{\frac{2k^3l^3}{\gamma_1^2+\epsilon_0^2}}e^{-k^2l^2/2}
\Bigl[1+\mathcal O\bigl((kl)^{-2}\bigr)\Bigr].
\label{E_n1_dle}
\end{equation}
The corresponding wave functions~(\ref{wavefun_hp_soln}) are the linear combinations of the bulk states $\Psi^{s\,(n=0)}_-$ and $\Psi^{s\,(n=1)}_-$ given by Eq.~(\ref{sol_inf_plane}), which at $\Delta=0$ reside solely on the $A$ sublattice, and the two orthogonal edge states residing on the $B$ sublattice, which at $y\ll y_k$ are given by
\begin{equation}
\Psi^{s\,(\text{edge I})}_-(y,k)=
C_{\text I}\,e^{-ky+\frac{y^2}{2l^2}}
\begin{pmatrix}
1 \\
0 \\
0 \\
0
\end{pmatrix}
,
\end{equation}
\begin{equation}
\Psi^{s\,(\text{edge II})}_-(y,k)
=C_{\text{II}}\,e^{-ky+\frac{y^2}{2l^2}}
\begin{pmatrix}
c-2ky \\
0 \\
-\frac{2\hbar v_Fk}{\gamma_1} \\
0
\end{pmatrix}
,
\end{equation}
where $c=1+(kl)^{-2}+\mathcal O\bigl((kl)^{-4}\bigr)$ is independent of $y$. More specifically, the wave functions corresponding to the lower energy solutions~(\ref{E_n0_sle}) in the main order in $1/(kl)$ are given by the mix of $n=0$ bulk state and the single-layer edge state,
\begin{align}
&\Psi^{s\,(a)\pm}_-\simeq\frac1{\sqrt2}\Bigl[\Psi^{s\,(n=0)}_-\pm\Psi^{s\,(\text{edge I})}_-\Bigr] \nonumber \\
&\quad+\frac1{2kl}\biggl[\sqrt{1+\epsilon_0^2/\gamma_1^2}\Psi^{s\,(n=1)}_-
\mp\frac{\epsilon_0}{\gamma_1}\Psi^{s\,(\text{edge II})}_-\biggr],
\end{align}
while the higher energy solutions~(\ref{E_n1_dle}) in the main order in $1/(kl)$ correspond to the hybridized bulk $n=1$ state and the double-layer edge state,
\begin{align}
&\Psi^{s\,(b)\pm}_-\simeq\frac1{\sqrt2}\Bigl[\Psi^{s\,(n=1)}_-\pm\Psi^{s\,(\text{edge II})}_-\Bigr] \nonumber \\
&\quad-\frac1{2kl}\biggl[\sqrt{1+\epsilon_0^2/\gamma_1^2}\Psi^{s\,(n=0)}_-
\mp\frac{\epsilon_0}{\gamma_1}\Psi^{s\,(\text{edge I})}_-\biggr].
\end{align}

\section{Limit of zero magnetic field}
\label{Appendix:zeroB}

In the limit $B\to0$, the argument $z$ of the parabolic cylinder function $U(a,z)$ in Eqs.~(\ref{f}), (\ref{hp_zigzag_disp_eq_plus}), and (\ref{hp_zigzag_disp_eq_minus}) is proportional to $l\to\infty$ while its complex parameter $a$ grows as $l^2$ (at fixed wave vector and energy). The appropriate asymptotic formula for this case is\cite{Olver1959JRNBS}
\begin{equation}
U\biggl(-\frac{\mu^2}2,\mu t\sqrt2\biggr)=g(\mu)\frac{e^{-\mu^2\xi(t)}}{(t^2-1)^{\frac14}}
\Bigl[1+\mathcal O\bigl(|\mu|^{-2}\bigr)\Bigr],
\label{main_asympt}
\end{equation}
where
\begin{equation}
\xi(t)=\frac12t\sqrt{t^2-1}-\frac12\ln\Bigl(t+\sqrt{t^2-1}\Bigr).
\label{xi}
\end{equation}
It is valid as $|\mu|\to\infty$, uniformly with respect to $t\in\mathbf S(\arg \mu)$ and $\arg\mu\in[-\pi+\epsilon,\pi-\epsilon]$, where $\epsilon$ is an arbitrary positive small constant. We will not need the explicit form of the function $g(\mu)$. The $t$ domain of validity $\mathbf S$, as well as the choice of the branch of the multivalued functions in Eqs.~(\ref{main_asympt}) and~(\ref{xi}), has a rather complicated dependence on $\arg\mu$ (see Ref.~\onlinecite{Olver1959JRNBS} for details). For our purposes, it is sufficient to know that the expansion is valid with the multivalued functions taken on their principal branches when $\mu t$ is real, $\re\sqrt{t^2-1}>0$ and $|\arg t|\ne\pi$ (the value at $t=0$ is obtained by continuity).

From~(\ref{main_asympt}) we get
\begin{equation}
\frac{U\bigl(-\mu^2/2,(\mu t+s/\mu)\sqrt2\bigr)}{U\bigl(-\mu^2/2,\mu t\sqrt2\bigr)}=e^{-s\sqrt{t^2-1}}
\Bigl[1+\mathcal O\bigl(|\mu|^{-2}\bigr)\Bigr],
\label{asympt_1}
\end{equation}
and for the derivative $U'(a,z)\equiv\partial U(a,z)/\partial z$ we obtain from~(\ref{main_asympt})
\begin{align}
U'\biggl(-\frac{\mu^2}2,{}&\mu t\sqrt2\biggr)=-\frac{\mu}{\sqrt2}g(\mu) \nonumber \\
&\times (t^2-1)^{\frac14}e^{-\mu^2\xi(t)}
\Bigl[1+\mathcal O\bigl(|\mu|^{-2}\bigr)\Bigr].
\label{derivative_asympt}
\end{align}
Using Eqs.~(\ref{main_asympt}) and~(\ref{derivative_asympt}) and recurrence relation~(\ref{recurrU}) for the parabolic cylinder function, one has
\begin{equation}
\frac{U\bigl(1-\mu^2/2,\mu t\sqrt2\bigr)}{U\bigl(-\mu^2/2,\mu t\sqrt2\bigr)}=\frac{\sqrt2}\mu\Bigl(t-\sqrt{t^2-1}\Bigr)
\Bigl[1+\mathcal O\bigl(|\mu|^{-2}\bigr)\Bigr].
\label{asympt_2}
\end{equation}
Taking $\mu=\sqrt{2\lambda}$, $t=-kl/\sqrt{2\lambda}$, and $s=\sqrt{2\lambda}\,y/l$, we obtain from Eqs.~(\ref{asympt_1}) and~(\ref{asympt_2})
\begin{equation}
\frac{U\bigl(-\lambda,\sqrt2(y/l-kl)\bigr)}{U\bigl(-\lambda,-\sqrt2kl\bigr)}=e^{-y\sqrt{k^2-2\lambda/l}}
\Bigl[1+\mathcal O\bigl(|\lambda|^{-1}\bigr)\Bigr],
\label{asympt_3}
\end{equation}
\begin{equation}
\frac{U\bigl(1-\lambda,-\sqrt2kl\bigr)}{U\bigl(-\lambda,-\sqrt2kl\bigr)}=
-\frac{kl+\sqrt{k^2l^2-2\lambda}}{\sqrt2\lambda}
\Bigl[1+\mathcal O\bigl(|\lambda|^{-1}\bigr)\Bigr],
\label{asympt_4}
\end{equation}
where $\re\sqrt{k^2l^2-2\lambda}>0$ and $k$ must be negative if $\lambda$ is real. Using Eq.~(\ref{asympt_4}) with $\lambda=\lambda^+_i-1/2$ and taking into account that in the limit $l\to\infty$
\begin{equation}
\sqrt{k^2-2\lambda^\xi_{1,2}/l}\to\kappa_\pm,
\label{lambdalimit}
\end{equation}
where $\kappa_\pm$ are defined in Eq.~(\ref{kappa}), the left-hand side of dispersion equation~(\ref{hp_zigzag_disp_eq_plus}) can be written in this limit as
\begin{align}
&U\bigl(1/2-\lambda_1^+,-\sqrt2kl\bigr)U\bigl(1/2-\lambda_2^+,-\sqrt2kl\bigr) \nonumber \\
&\times \frac{(E^2-\Delta^2)(\kappa_+-\kappa_-)}{\sqrt2\epsilon_0\gamma_1(\lambda_1-\lambda_2)}
\biggl[1-\frac{\epsilon_0^2l^2(k+\kappa_+)(k+\kappa_-)}{2(E-\Delta)^2}\biggr],
\end{align}
which immediately implies Eq.~(\ref{disp_eqn_zeroB}) with the upper sign for the left-edge subgap mode in the $K_+$ valley.

In the same way, we apply Eq.~(\ref{asympt_4}) with $\lambda=\lambda^-_i+1/2$ to the half-plane dispersion equation~(\ref{hp_zigzag_disp_eq_minus}), which leads to the equation
\begin{equation}
\hbar^2v_F^2(k-\kappa_+)(k-\kappa_-)=(E-\Delta)^2
\label{KminusB0dispEq}
\end{equation}
for the left-edge subgap mode in the $K_-$ valley.

The corresponding wave functions of the left-edge states in both valleys can be obtained from Eq.~(\ref{f}) by using Eqs.~(\ref{asympt_3})--(\ref{lambdalimit}). The result reads
\begin{align}
&\Psi^s_\pm(y,k)=
\sum_{\sigma=\pm}C^\pm_\sigma e^{-\kappa_\sigma y} \nonumber \\
&\;\;\times
\begin{pmatrix}
-1 \\
\mp\frac{E\mp\Delta}{\hbar v_F(k+\kappa_\sigma)} \\
\pm\frac{\hbar^2 v_F^2(k^2-\kappa^2_\sigma)-(E\mp\Delta)^2}{\hbar v_F\gamma_1(k+\kappa_\sigma)} \\
\frac{(E\mp\Delta)(\gamma_1^2-E^2+\Delta^2)+\hbar^2 v_F^2(k^2-\kappa^2_\sigma)(E\pm\Delta)}
{\hbar^2v_F^2\gamma_1(k+\kappa_\sigma)^2}
\end{pmatrix}
,
\label{B0wavefun}
\end{align}
in agreement with Ref.~\onlinecite{Li2010PRB}.

The zero-energy solution for the left-edge subgap state in both valleys $K_{\xi=\pm}$ at $B=0$ can be found analytically. This state is located at $\hbar v_Fk=-\xi\gamma_1$,\cite{Li2011NP} and the corresponding wave function~(\ref{B0wavefun}) is given by
\begin{align}
\Psi^s_{\xi A_i}(y)&=Ce^{-\frac{\gamma_1y}{2\hbar v_F}}\sin\biggl(\frac{\Delta y}{\hbar v_F}\biggr), \quad i=1,2, 
\nonumber \\
\Psi^s_{\xi B_1}(y)&=-Ce^{-\frac{\gamma_1y}{2\hbar v_F}}\cos\biggl(\frac{\Delta y}{\hbar v_F}\biggr),
\label{E0B0wavefun} \\
\Psi^s_{\xi B_2}(y)&=-\Psi^s_{\xi B_1}(y)-\frac{\gamma_1}\Delta\Psi^s_{\xi A_1}(y). \nonumber
\end{align}

The dispersion equation of the right-edge subgap mode in the $K_+$ valley [Eq.~(\ref{disp_eqn_zeroB}) with the lower sign] and the corresponding wave functions are obtained from Eqs.~(\ref{KminusB0dispEq})--(\ref{E0B0wavefun}), using the correspondence~(\ref{zigzag_E_symmetry}) and~(\ref{zigzag_psi_symmetry}) between the spectra and the wave functions in two valleys.

\end{document}